\documentclass[epj]{svjour}
\usepackage{graphicx}
\usepackage{bm}   
\usepackage{color}
\usepackage{amssymb}
\usepackage{amsmath}
\usepackage{balance}
\usepackage{subfigure}

\begin{document}
	\title{Extreme events in the Li\'enard system with asymmetric potential: An in-depth exploration}
	\author{B. Kaviya\and R. Suresh \and V. K. Chandrasekar \mail{suresh@eee.sastra.edu}
	}                     
	%
	%
	\institute{Department of Physics, Centre for Nonlinear Science and Engineering, School of Electrical and Electronics Engineering,\\ SASTRA Deemed University, Thanjavur 613 401, India.}
	\date{Received: -- / Revised version: --}
	%
	\abstract{
This research investigates the dynamics of a forced Li\'enard oscillator featuring asymmetric potential wells. We provide compelling evidence of extreme events (EE) in the system by manipulating the height of the potential well. In the case of a symmetric well, the system exhibits chaotic behavior, with the trajectory irregularly traversing between the two wells, resulting in frequent large oscillations under specific parameter values. However, the introduction of asymmetry in the potential wells induces a noteworthy transformation. The frequency of jumping between wells is significantly diminished. In essence, the system trajectory displays rare yet recurrent hops to the adjacent well, which we identify as EE. The intricate dynamical behavior observed in the system is elucidated through bifurcation diagrams and Lyapunov exponents. The emergence of EE in the system, governed by various parameters, is characterized using the threshold height, probability distribution function, and inter-event intervals. We illustrate the regions of EE using phase diagram plots and demonstrate the control of EE by incorporating a damping term into the system.
		\PACS{
			{PACS-key}{Li\'enard oscillator,  Asymmetric potential well,  Extreme events  }
		} 
	}
	\titlerunning{Extreme events in the Li\'enard system with asymmetric potential: An in-depth exploration}
	\authorrunning{B. Kaviya et al.}
	\maketitle

	\section{Introduction}
	\label{sec1}
	The signature of extreme events (EE) is ubiquitous and imprinted in both natural and engineering systems \cite{Albeverio-1,McPhillips-2,Akhmediev-3}. For example, EE are rare and observed in various forms such as floods \cite{Sachs-4}, cyclones \cite{Emanuel-5}, tsunamis \cite{Mascarenhas-6,Bird-7}, tornadoes, earthquakes, droughts, rainfall \cite{Ghil-8}, toxic algal blooms \cite{Anderson-9}, solar flares \cite{Buzulukova-10}, mass frenzies \cite{Helbing-11}, terrorist attacks \cite{Kunreuther-12}, industrial accidents \cite{Zio-13,Salzano-14}, share market crashes \cite{Johansen-15,Krause-16}, and power blackouts \cite{Dobson-17,Kinney-18} are few examples of EE. Additionally, EE have also been observed in many engineering models and experimental studies such as super fluid helium, laser systems, optical fiber, climatic studies, and EEG studies on epileptic conditions in rodents \cite{Reinoso-19,Solli-20,Bailung-21,Ganshin-22,Pisarchik-23}. Due to the universality of occurrence and deleterious nature of EE excites the research community from distinct disciplines to explore this behavior in various aspects. 
	
A major challenge in studying EE lies in the scarcity of experimental data. In many cases, acquiring data on EE is not only difficult but sometimes entirely unfeasible. In such instances, dynamical models provide a critical alternative, allowing researchers to systematically manipulate parameters and replicate conditions that resemble real-world scenarios. This ability to control and fine-tune parameters offers unique insights into the mechanisms that drive EE. As a result, dynamical models have become instrumental in investigating EE across a wide range of systems. These include well-known models such as the Ginzburg-Landau equation, the nonlinear Schrödinger equation, Liénard systems, and others \cite{Ansmann-24,Reinoso-25,Suresh-26,Suresh2018,Kingston-27,kaviya-47,Karnatak-28,Saha-29,Saha-30,Rothkegel-31,Ray-32,Ray-33}. Over the past few decades, these models have revealed numerous underlying mechanisms responsible for the emergence of EE \cite{chowdhury-46}, advancing our understanding of these rare but significant phenomena. These includes interior interior crisis\cite{Grebogi-34,Ditto-35}, PM intermittency\cite{Pomeau-37,Jeffries-38}, On-off intermittency\cite{Xie-42}, in-out intermittency\cite{Ashwin-43}, sliding bifurcation near a discontinuous boundary\cite{Kumarasamy-36}, instability in the synchronization manifold\cite{Mishra-39,Mendoza-40}, Pulse-shaped explosion\cite{kaviya1-41}, SNA breakdown\cite{kaviya_sna-40}, interaction of a chaotic attractor with the saddle orbit\cite{Ray-51}, quasiperiodic route\cite{Mishra-44,Nicolis-45}, etc. These diverse roots and mechanisms have been identified as potential catalysts for EE in dynamical systems, whether they are isolated, coupled, or as a part of network motifs. While significant progress have been made in understanding and predicting EE, the complexity of dynamical systems means that there are still many interesting problems to be explored in this domain. 
	\begin{figure}[h!]
		\centering
		\includegraphics[width=0.6\columnwidth]{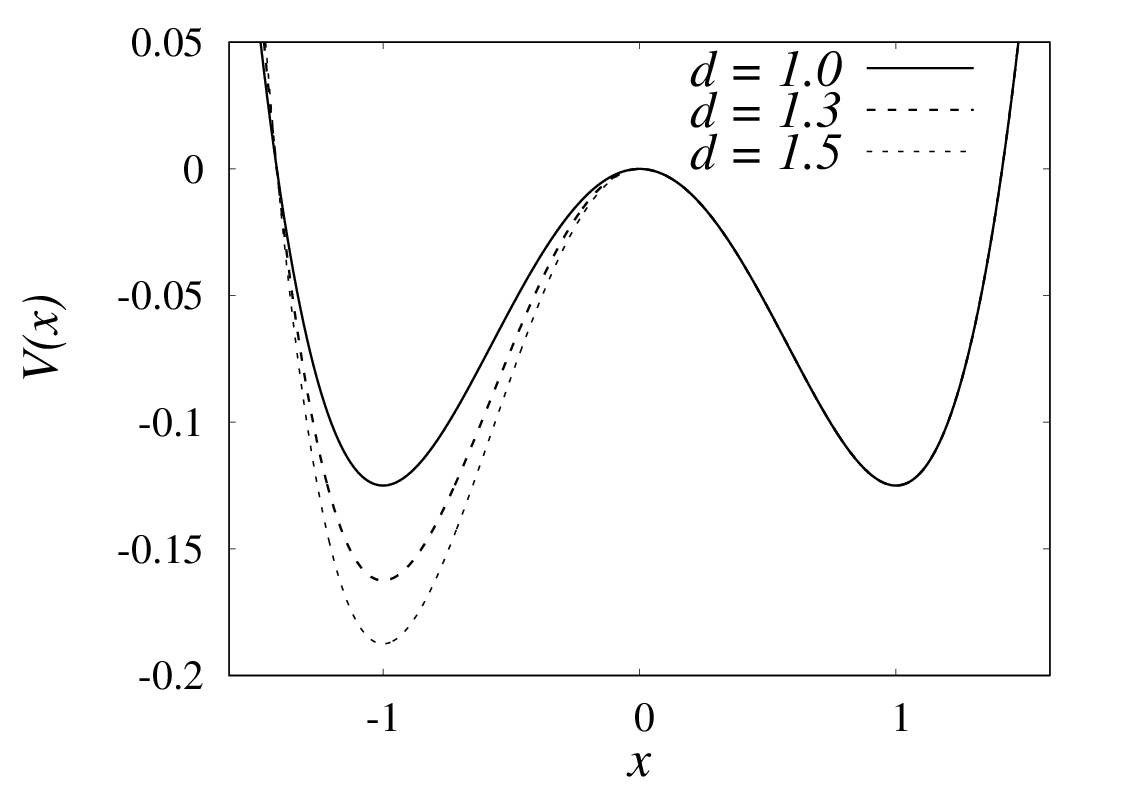}
		\caption{The form of the potential ($V$) within the left well can be altered by varying the  critical parameter $d$ = 1.0 (continuous line), $d$ = 1.3 (dashed line), and $d$ =  1.5 (dotted line)}
		\label{fig1}
	\end{figure}
	One such challenging yet interesting problem is the study of EE in dynamical systems with asymmetric potential well. This type of asymmetric potential scenario has been extensively explored in various angles, including the study of bursting oscillations, resonance phenomena, and energy harvesting perspectives across a wide class of nonlinear systems \cite{simo2016,ravichandran2009,litak2022}. For instance, investigations have been conducted into the emergence of bursting oscillations in duffing oscillator when subjected to three different forms of asymmetrical double-well potential \cite{simo2016}. Additionally, research has delved into the chaotic dynamics exhibited by the duffing oscillator when subjected to an asymmetric double-well potential \cite{ravichandran2009}. Further, stochastic resonance characteristics of a duffing oscillator with asymmetric potential well have been carried out by S. Arathi et al \cite{aarthi-48}. Their findings indicate that the degree of asymmetry has a significant impact on several key factors, such as signal to noise ratio and average residence time of the trajectory both in left and right potential wells during resonance. Additionally, Gammaitoni et al. introduced a continuous bias into the system to explore the emergence of stochastic resonance in an over-damped asymmetric double well system \cite{Gammaitoni-49}. Further, recent research by Borromeo and Marchesoni reported the emergence of double stochastic resonance while studying the dynamics of an asymmetric system\cite{borromeo-52}. Furthermore, in ref \cite{liu2022}, the authors investigates the conditions for the occurrence of meta-stable chaos in a bi-stable asymmetric laminated composite shallow shell subjected to different excitation.  However, despite this extensive research, no previous studies have been reported the advent of EE in the context of systems with asymmetric potential wells. 
	
	Therefore, we posted the challenge of exploring dynamics of a Li\'enard oscillator, which is basically a damped and driven anharmonic oscillator, influenced by varying asymmetric potential heights. When the potential well is symmetric, and for a chosen parameter values, the system exhibits chaotic behavior, with trajectories unpredictably transitioning between the two wells, resulting in frequent large oscillations. Whereas, when we alter the height of the potential on the left side of the well. We find that this height variation induces a repulsive effect on approaching trajectories, via unstable fixed point, resulting in rare crossing of trajectories into the left well. On the other hand, notably, this behavior contradicts what happens on the right side of the well, where the stable equilibrium resides, causing the trajectories to converge around the stable fixed point. It means that the frequency of hopping between wells is significantly reduced and made rare but recurrent jumping to the left potential well, which we identify as EE. This alternation in behavior is characterized using bifurcation diagrams, and various dynamical states are confirmed by Lyapunov exponents (LEs)~\cite{Benettin,Lak}, revealing the system's rich dynamics nature. Further, we use threshold height to characterize EE, and the long tail nature of EE is confirmed by plotting probability distribution function (PDF). Furthermore, the controlling mechanism of EE is attempted by incorporating a linear damping term into the system. 
	
	\begin{figure}[h!]
		\centering
		\includegraphics[width=1.0\columnwidth]{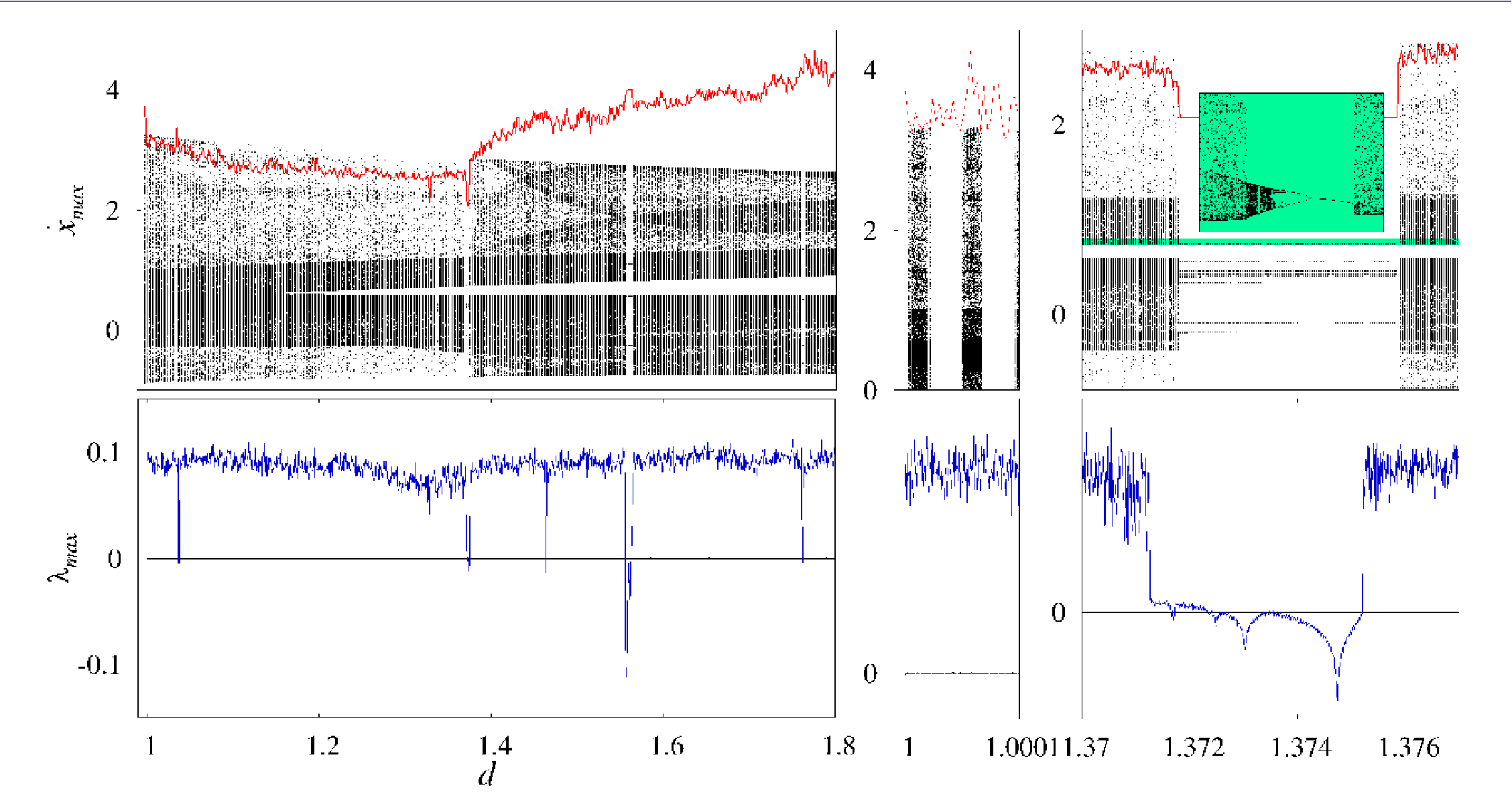}
		\caption{The top-left panel presents the bifurcation diagram for system (\ref{Eqn1}), with $\xi$ = 0, plotted against the potential height ($d$). It illustrates the qualitative changes observed in the maxima of the $\dot{x}$, denoted as $\dot{x}_{max}$, while maintaining constant values for the other parameters: $\omega=0.6460$, $\alpha=0.45$, $\beta=0.5$, $\gamma=0.5$ and $f=0.2$. The top-middle and top-right panels serve as insets of the bifurcation diagram. The bottom-panel depicts the Lyapunov exponent spectrum, confirming the emergence of both regular and chaotic dynamics as a function of $d$.}
		\label{fig2}
	\end{figure}
	\begin{figure}[h!]
		\centering
		\includegraphics[width=0.6\columnwidth]{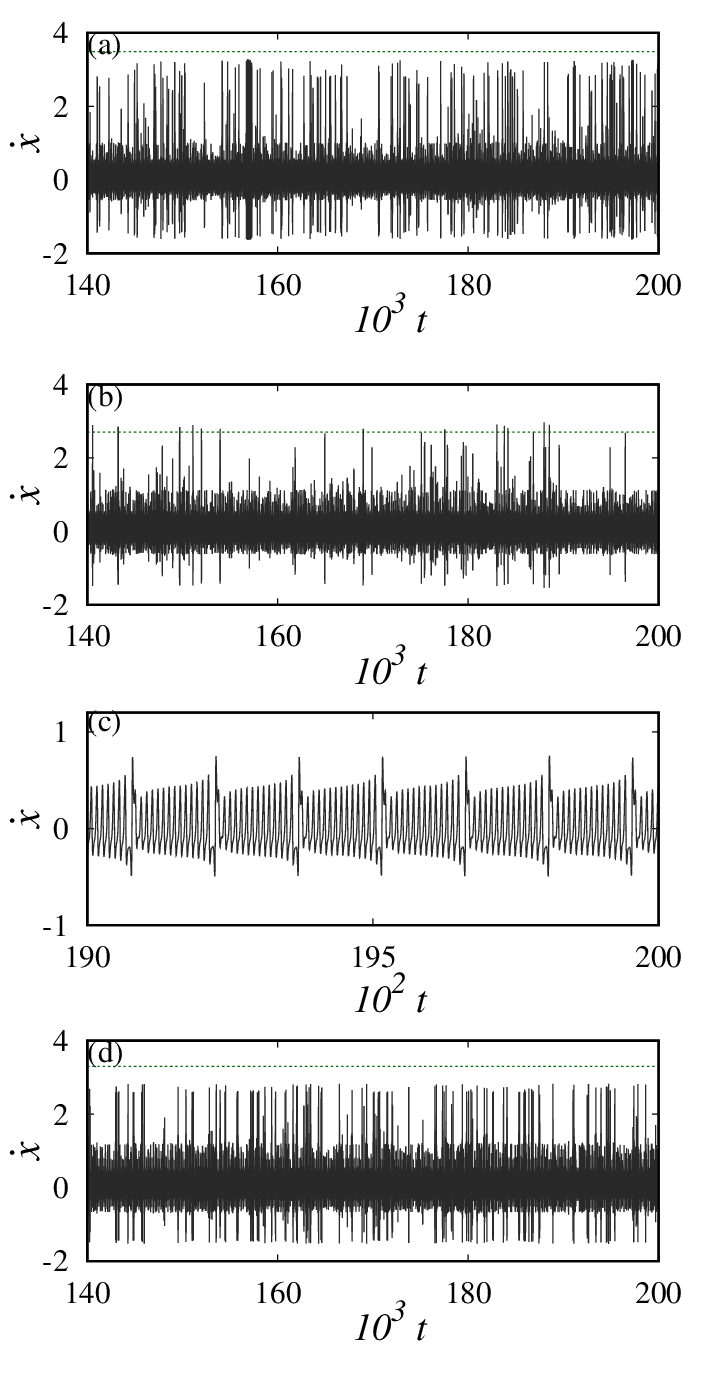}
		\caption{Time series plots are provided for various potential heights ($d$). In particular, (a) shows the frequent occurrence of large-amplitude oscillations, corresponding to a potential height of $d=1.0$ ($H_T$=3.3521), (b) display the emergence of rare but recurring large peaks, denoting extreme events, correspond to $d=1.1$  ($H_T$=2.692), (c) illustrates regular oscillations for $d=1.374$. Finally, (d) once again show the frequent emergence of large-amplitude oscillations for $d=1.5$   ($H_T$=3.64). The horizontal dotted line denotes the threshold height ($H_T$).}
		\label{fig3}
	\end{figure}
	 The subsequent sections of this article are structured as follows: Section \ref{sec2} presents the chosen mathematical model for our study and elucidates the emergence of EE. Section \ref{sec3} focuses on the impact of linear damping on the Li\'enard system and outlines strategies for EE control. Finally, Section \ref{sec4} provides a synthesis of our findings along with concluding remarks.
	\begin{figure}[h!]
		\centering
		\includegraphics[width=1.0\columnwidth]{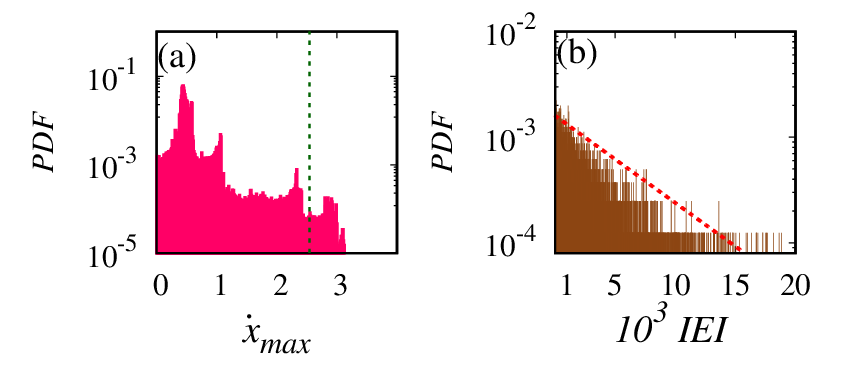}
		\caption{The probability distribution function (PDF) reveals the presence of long tail behavior, providing confirmation for the occurrence of EE when $d$ = 1.1. The remaining parameters are fixed as given in Fig. \ref{fig1}. The vertical dashed line is employed to represent the threshold height ($H_T$). The IEI distribution histogram follows the Poisson fit (plotted in a red dotted line) confirming the rare occurrence of EE for $d=1.1$. The remaining parameters are fixed as given in Fig. \ref{fig1}}
		\label{fig4}
	\end{figure}
	\section{Dynamics of the forced Li\'enard oscillator with asymmetric potential}
	\label{sec2}
	 To demonstrate our findings, we utilized a straightforward yet multistable Li\'enard system~\cite{Kingston-2017,chan}, which serves as a representative model for a broad class of parity ($\cal{P}$) and time ($\cal{T}$) reversal symmetric ($\cal{PT}$-symmetric) systems for a particular parameter choice. This Li\'enard type systems have widespread applications across various domains, such as optics~\cite{markis}, solid-state physics~\cite{Joglekar}, metamaterials~\cite{Lazarides}, optomechanical systems~\cite{Jing}, and more. The detailed study of the Li\'enard system and its dynamics, particularly concerning the generation of limit cycle oscillations and harmonic solutions, has been extensively explored in existing literature\cite{Sanjuan1998,ruiz2000,zhou2023,giacomini1998,adjai2022,chan}. In this study, we developed a mathematical model of the Li\'enard system, where we manipulated the height of one of the potential wells by adjusting a specific parameter. The system's equation of motion is described by
	\begin{eqnarray}
		\dot{x}&=& y, \nonumber \\
		\dot{y}&=& -\alpha xy-\frac {dV}{dx}-\xi y+f\sin(\omega t).
		\label{Eqn1}
	\end{eqnarray} 
	In this Eq.(\ref{Eqn1}), we define the parameters as follows: $\alpha$ and $\xi$ denotes the nonlinear, and linear damping coefficients, respectively, while $f$ and $\omega$ correspond to the amplitude and frequency of the external periodic forcing, respectively. The autonomous Li\'enard-type equation plays a crucial role in describing phenomena such as the spherically symmetric expansion and collapse of a gravitating mass\cite{mcvittie1933}, the behavior of a spherical gas cloud under molecular interaction and thermodynamic laws \cite{leach1985,dixon1990}, and the modeling of fusion processes in pellets \cite{erwin2006}. 
		
The Li\'enard system exhibits a distinctive dual nature, acting as a cubic anharmonic oscillator with a nonstandard Hamiltonian and a conservative nonlinear oscillator influenced by a nonlinear damping term $(\alpha xy)$\cite{chan}. In this context, the nonlinear damping term serves a dual role, acting as both a damping and a pumping term, thereby modulating the occurrence of self-sustained oscillations based on the oscillation amplitudes~\cite{Chandrasekar,mishra-2015}. This type of oscillatory model, featuring similar nonlinear damping, is prevalent across a spectrum of physical, mechanical, and chemical systems, as well as engineering models, particularly when appropriate transformations are applied. For instance, nonlinear damping coefficients resembling restoring forces or stiffness coefficients are commonly observed in micro and nanoelectromechanical systems\cite{Eichler}, mirroring the nonlinear elements frequently encountered in electronic circuits. Moreover, the autonomous Li\'enard oscillator exhibits bistability, where trajectories either converge to stable fixed points or display self-sustained periodic oscillations, contingent upon initial conditions.

The derivative of the potential function, $\frac{dV}{dx}$, in Eqn. (\ref{Eqn1}) plays a critical role in defining the restoring force within the system, directly influencing its dynamical behavior. In this study, the asymmetric potential $V(x)$ creates disparities in the stability of the two wells, significantly affecting phase-space trajectories and the occurrence of EE.

For  $x\ge0$, the potential is symmetric, resulting in equal restoring forces in both wells. This symmetry leads to frequent transitions between the wells, accompanied by chaotic oscillations. However, for $x<0$, the asymmetry introduced by the parameter $d$ causes a steeper slope in one well and a shallower slope in the other. As a result, the derivative  $\frac{dV}{dx}$ is smaller in the shallower well, making transitions between wells more difficult. The reduced slope creates a repulsive effect near the unstable fixed point in the left well, lowering the frequency of transitions. Consequently, the system tends to remain confined within one well for extended periods, leading to rare but recurrent extreme oscillations when transitions do occur--characteristic of EE.
	\begin{figure}[h!]
	\includegraphics[width=0.9\columnwidth]{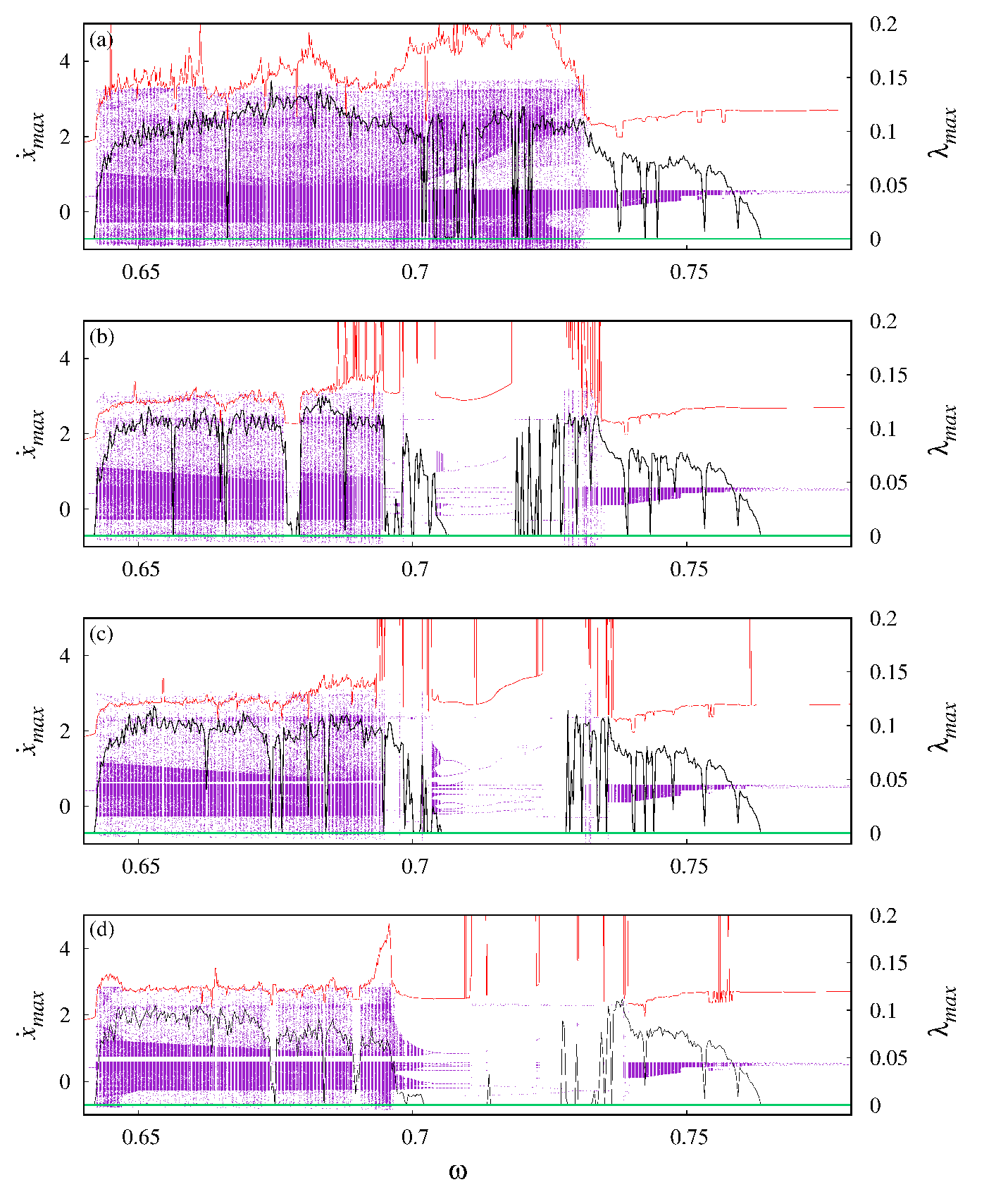}
	\caption{Bifurcation diagrams, accompanied by their respective LE spectra (indicated by the black line), are plotted as function of the external forcing frequency ($\omega$) for system (\ref{Eqn1}), with $\xi$ = 0,  shows the qualitative changes that occurred in the maxima of the $\dot{x}$ variable denoted as $\dot{x}_{max}$ across various potential heights ($d$). (a) corresponds to $d=1.0$, (b) corresponds to $d=1.1$, (c) corresponds to $d=1.3$, and (d) is plotted for $d=1.5$. The red (dark gray) line in each diagram represents the threshold height $H_T$, indicating the regions where EE occur. The green line represents the zero axis of the LE, aiding in distinguishing chaos from the periodic region.}
	\label{fig6}
\end{figure}

The system’s dynamics are thus governed by the interplay between the asymmetry of the potential and the derivative $\frac{dV}{dx}$. The depth of the wells, modulated by the parameter $d$, influences the strength of the restoring force and the likelihood of extreme transitions. Increasing the asymmetry further reduces transition probabilities, enhancing both the rarity and extremity of events when they occur.

\noindent The expression for potential $V(x)$ for Eqn. (\ref{Eqn1}) is given as
	\[
	V(x) = 
	\left\{\begin{array}{ccr}
		-\frac{1}{2}\gamma x^{2}+\frac{1}{4}\beta x^{4} & x\geq0 \\	-\frac{1}{2}\gamma dx^{2}+\frac{1}{4}\beta dx^{4} & x<0.
	\end{array}\right.
	\]
	Here parameter $d$ controls the height of the well. We have modified the height of the left potential well by adjusting the parameter $d$. One can visualize the asymmetrical potential of Li\'enard oscillator with different potential heights for $d$ values of 1, 1.3, and 1.5 in the Fig. \ref{fig1}. 
	
	Initially, we investigate the dynamics of the Li\'enard system (\ref{Eqn1}) without the presence of linear damping by fixing $\xi$ = 0. The equations were numerically solved using the fourth-order Runge-Kutta method with a step size of 0.01. To facilitate the numerical simulations, we have fixed the system parameters constant at $\alpha$ = 0.45, $\beta$ = $\gamma$ = 0.5, and $f$ = 0.2. We then systematically  vary the parameters $\omega$, and $d$ in order to examine the system's dynamics. 
	
	We observed a range of dynamical states in the system while varying the parameters $d$, and $\omega$. To illustrate this, we initially computed the bifurcation diagram and the corresponding maximum positive LE spectrum of the system (\ref{Eqn1}) as a function of the parameter $d$ while keeping the frequency of external forcing $\omega$ at 0.646. The upper left panel of Fig. \ref{fig2} illustrates the qualitative changes that observed in the maxima of the $\dot{x}$ variable ($\dot{x}_{max}$), while the corresponding LE spectrum is depicted in the lower left panel.  Both analyses were conducted by simulating Eq. (\ref{Eqn1}) over a period of $5\times10^{6}$ time units, with a sufficient transition period allowed for the system to reach a steady state before data collection began. When $d$ is set to 1, the two potential wells have equal heights resulting in a symmetric potential well configuration. In this case, the system exhibits chaotic behavior with positive LE for the chosen parameters. The trajectories of the system are irregularly hop between the two potential wells, resulting in frequent large-amplitude peaks amidst small-amplitude oscillations. We captured the time evolution of this scenario after excluding the large transient dynamics and plotted it in Fig. \ref{fig3}(a) for $d$ = 1, showing the frequent occurrence of high peaks in the $\dot{x}$-variable of the system (\ref{Eqn1}) as a function of time $t$. 
	
	As soon as the parameter $d$ increased beyond 1, the potential well became asymmetric while the system continued to exhibit chaotic dynamics but with some differences. Significantly, the jumping of trajectories to the left potential well was considerably reduced. Consequently, the system displayed infrequent yet recurrent high-amplitude oscillations, referred to as EE. Among those high peaks, only a few exceeding the pre-defined threshold height denoted as $H_T$, calculated as $\left\langle\dot{x}_{max}\right\rangle+7\sigma$, where $\left\langle\dot{x}_{max}\right\rangle$ represents the time averaged maxima of the $\dot{x}$-variable, and $\sigma$ is the mean standard deviation. In simpler terms, peaks exceeding 7 times the standard deviation were identified as EE. We determined the threshold height for Eq. (\ref{Eqn1}) by analyzing a large datasets, excluding sufficient transitions. 
	
	In the bifurcation diagrams presented in Fig. \ref{fig2}, the threshold height, $H_T$ is delineated by a red line.  Meanwhile in Fig. \ref{fig3}, horizontal dotted lines indicate this threshold height. The temporal evolution of the dynamical state for $d$ = 1.1 is depicted in Fig. \ref{fig3}(b) showcasing relatively infrequent high peaks in comparison to the scenario with a symmetric potential well [Fig. \ref{fig2}(a)].  In Fig. \ref{fig3}(a), all the large peaks fall below the $H_T$ line, confirms the absence of EE in the system when $d$ = 1. This observation is further supported by the mid-upper panel of Fig. \ref{fig2}, where all values of $\dot{x}_{max}$ remain below the $H_T$ line for $d=1$. Conversely, for $d$ = 1.1 [Fig. \ref{fig3}(b)], a few peaks exceed the $H_T$ threshold, corroborating the emergence of EE. 
	
	The long tail behavior, which typically indicates the presence of EE, is often associated with heavy-tailed distributions. In such distributions, the probability of observing EE decreases more slowly as the amplitude of the events increases, in contrast to a normal (Gaussian) distribution. To access the long-tail behavior of the system's $\dot{x}_{max}$ variable, we estimate its PDF in semi-log scale, as illustrated in Figure \ref{fig4}(a). This representation clearly illustrates that the system exhibits events exceeding the predefined threshold height $H_T$ (indicated by the dashed vertical line). Consequently, providing further confirmation of the presence of EE at $d$ = 1.1. Another crucial characteristic of EE is their infrequent occurrence, a feature we verified by analyzing the inter-event interval (IEI) histogram. This analysis involved computing the IEI using the formula $IEI_n = t_{n+1} - t_n$, where $t_n$ represents the occurrence time of the $n^{th}$ event among the total number ($n$) of EE in the system. In Fig. \ref{fig4}(b), we illustrate the estimated IEI histogram for $d = 1.1$, showing that the probability of a short IEI occurrence is much higher than that of a longer IEI occurrence, corroborating the rarity of these events with a Poisson fit. The red (dark gray) dotted line in the figure corresponds to the Poisson distribution, estimated with a slope value of 0.0009005. Notably, to evaluate the PDF and IEI, we generated extensive datasets of time series comprising $5 \times 10^8$ data points, ensuring sufficient representation and statistical significance.
	
	Continuing to increase the parameter $d$, the system exhibits EE until reaching approximately $d\approx$ 1.37. Beyond this point, large-amplitude oscillations, or EE, suddenly vanish from the system's dynamics, and the system exhibits small-amplitude chaos within a narrow range of $d\in[1.371,1.373)$. In other words, when we decrease the parameter $d$ from higher to lower values (moving from right to left in the bifurcation diagram), the small amplitude chaotic attractor is suddenly transforms into a larger chaotic attractor via an interior crisis \cite{Grebogi-34,Ditto-35}. This transformation is clearly visible in the inset of the bifurcation diagram, which is plotted in the upper right panel of Fig. \ref{fig2}.  Continue to increase $d$  eventually leads the system into periodic dynamics via period-adding bifurcation route, as evident from the bifurcation diagram. The negative values of maximum LE displayed in the lower right panel support the presence of periodic dynamics. The time evolution of the system for $d$ = 1.374 is illustrated in Fig. \ref{fig3}(c), revealing  mixed-mode oscillations like multi-periodic dynamics.  It is worth noting that  similar periodic windows intermittently appear in the bifurcation diagram for various values of $d$.  
	
	Subsequently, an examination of the dynamics of the system (\ref{Eqn1}) without linear damping ($\xi$ = 0) is undertaken by systematically varying the external forcing frequency ($\omega$). Specifically, the computation of the bifurcation of the maxima of the $\dot{x}$-variable ($\dot{x}_{max}$) against $\omega\in(0.64, 0.757)$ is conducted for different values of $d$. To identify various dynamical states and characterize EE, maximal LE and $H_T$ are calculated and plotted. In Figure \ref{fig6}(a), the bifurcation dynamics are illustrated for the case of a symmetric potential well ($d$ = 1). Notably, large-amplitude oscillations (or EE) manifest through two distinct bifurcation routes as a function of $\omega$ \cite{Suresh2018}. An intermittency route is observed as $\omega$ increases from lower to higher values (as depicted in Fig. \ref{fig6}(a) from left to right), while the interior crisis route is evident when $\omega$ decreases from higher to lower values (as shown in Fig. \ref{fig6}(a) from right to left). It is noteworthy that within the $\omega\in(0.6423, 0.7316)$ range, the system exhibits large chaotic oscillations (both EE and non-EE) with the emergence of a narrow range of intermittent periodic windows, as evident from Fig. \ref{fig6}(a).
	
	Upon introducing asymmetry in the potential well and increasing the asymmetry parameter $d$, intriguing bifurcations occur in the system. Specifically, the region of EE emerging via the intermittency route widens in the $\omega$ parameter space as a function of the critical parameter $d$. Subsequently, the EE region slowly diminishes in the $\omega$ parameter region, and the frequency of large-amplitude oscillations increases, resulting in large amplitude double-well chaotic oscillations. This scenario is depicted for three different values of $d$ = 1.1, 1.3, and 1.5 in Figs. \ref{fig6}(b), \ref{fig6}(c), and \ref{fig6}(d), respectively. Notably, in comparison to Fig. \ref{fig6}(b), the region of $\omega$ to which the maxima of the $\dot{x}$-variable cross the threshold height $H_T$ is widened in Fig. \ref{fig6}(c). However, in Fig. \ref{fig6}(c), it is depicted that the $H_T$ line is larger than the $\dot{x}$-variable in most regions of $\omega$.
	
	Conversely, EE emerging via the interior crisis route while decreasing $\omega$ from large to small values is consistently reduced with an increase in $d$, as evident from Fig. \ref{fig6}. Furthermore, a comparison of Figs. \ref{fig6}(b), \ref{fig6}(c), and \ref{fig6}(d) with Fig. \ref{fig6}(a) reveals that the chaotic region in the $\omega$ parameter space is significantly reduced, and intermittent periodic windows emerge via period doubling and intermittency routes. This change in bifurcations can be identified from Figs. \ref{fig6}(b), \ref{fig6}(c), and \ref{fig6}(d).
	
\begin{figure}[h!]
		\begin{center}
			\includegraphics[width=1.0\textwidth]{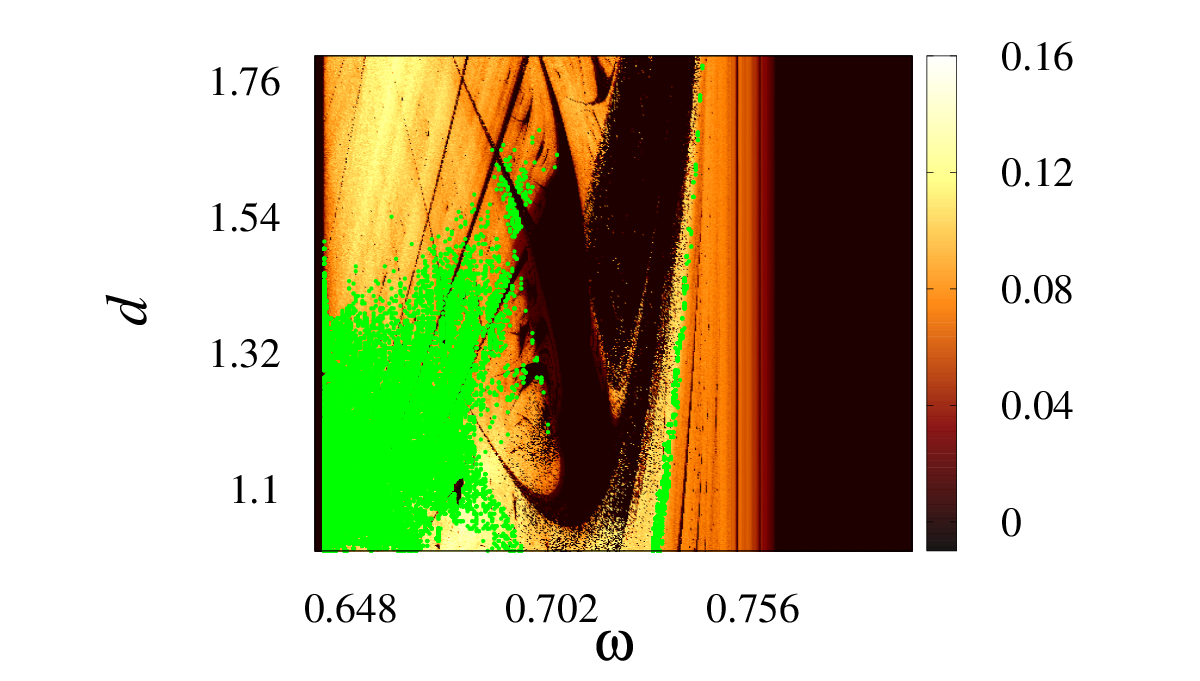}
		\end{center}
		\caption{Two parameter phase diagram of the system (\ref{Eqn1}), with $\xi$ = 0, as a function of $\omega\in(0.64, 0.757)$ and $d\in(1.0,1.8)$ illustrates the emergence of different dynamical states. The color gradient within the diagram corresponds to the LE values. Dark brown (gray) regions signify shades negative values in maximum LE values, confirming the regular dynamics of the system. Conversely, positive LE, depicted in remaining color shades, indicate chaotic dynamics. Further, the light green (gray) points within the diagram highlight the parameter values where occurrence of EE has been observed.}
		\label{fig5}
	\end{figure}

	The dynamical behavior of the system reveals a rich variety of states, encompassing bifurcations and transitions between different regimes, particularly as we altered the depth of the potential well. To extensively investigate and visualize these transitions across a wide spectrum of parameter values, we have carefully estimated a two-parameter phase diagram for Eq. (\ref{Eqn1}) with $\xi=0$. In this phase diagram, we utilize maximum LE to distinguish between chaotic and periodic regions, providing valuable insights into the system's behavior. Additionally, we apply a threshold height criterion to locate the parameters associated with the occurrence of EEs. The two-parameter phase diagram, illustrated in Figure \ref{fig5}, serves as a visual representation of the system's dynamic landscape, showcasing the emergence of various dynamical states. The color gradient within the diagram corresponds to the LE values. Specifically, regions colored in dark brown (or black) indicate the presence of negative values in positive LE, confirming the periodic dynamics within those parameter combinations. Conversely, the remaining color (or gray) shades represent positive LE, signifying chaotic dynamics. Notably, the light green (or gray) points scattered throughout the diagram highlight specific parameter values where the occurrence of EEs has been observed. This comprehensive phase diagram allows for an understanding of the system's complex behavior and its response to variations in $\omega$ and $d$.
	
	\section{Influence of linear damping on the forced Li\'enard oscillator with asymmetric potential}	\label{sec3}
	 This study establishes the manifestation of EE in the Li\'enard system featuring an asymmetric potential. Previous literature underscores the pivotal role of linear damping in governing intricate dynamics within nonlinear systems, as substantiated by prior research \cite{kaviya-47,kaviya1-41,kaviya_sna-40}. Motivated by this, our investigation focuses on examining the impact of linear damping on EE within the considered system.
		
\begin{figure}[h!]
		\includegraphics[width=1.0\columnwidth]{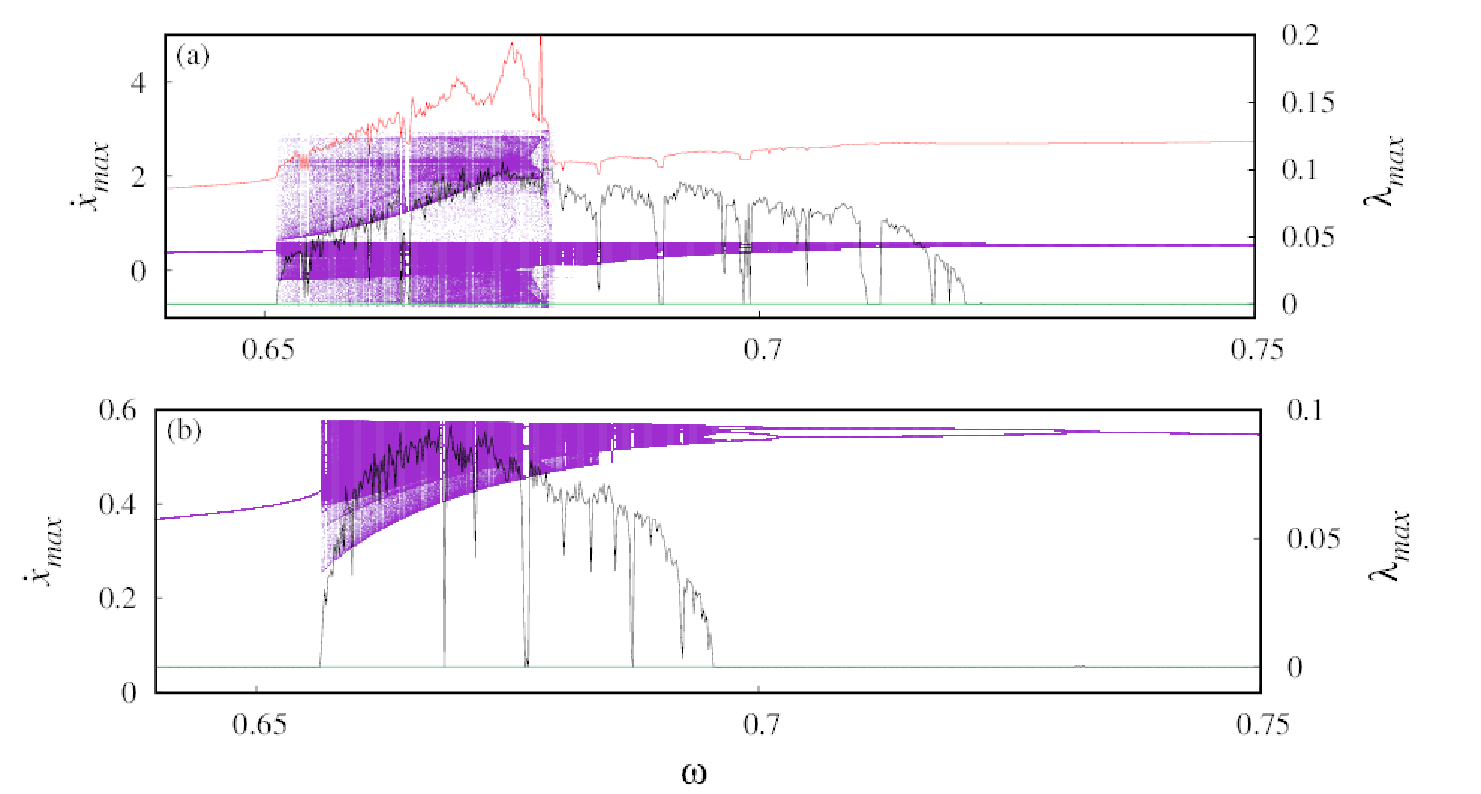}
		\caption{The bifurcation diagrams, accompanied by their respective Lyapunov exponent ($\lambda_{max}$) spectra (indicated by the black line), are plotted as functions of the external forcing frequency ($\omega$) for system (\ref{Eqn1}) with $\xi>0$. The plots depict the qualitative changes observed in the maxima of the $\dot{x}$ variable for $d=1.1$, considering two different values of $\xi$: (a) for $\xi=0.02$,  and (b) for $\xi=0.03$. The red (dark gray) line represents the threshold height $H_T$, delineating the regions where EE occur. Additionally, the green (light gray) line denotes the zero axis of the LE, aiding in the differentiation between chaotic and periodic regions.}
		\label{fig8}
	\end{figure}

		We begin by conducting bifurcation analysis and computing LE spectrum similar to Fig. \ref{fig6} for two different values of the linear damping parameter $\xi$, with the depth of the potential well fixed at $d=1.1$. Figure \ref{fig8}(a) illustrates the dynamics of the Li\'enard system for $\xi=0.02$, indicating a reduction in the region of EE in the $\omega$ parameter space compared to Fig. \ref{fig6}(b) (plotted for $\xi=0$). Additionally, the region of chaos diminishes as a function of $\omega$, as evident from the LE spectrum. Increasing the damping parameter further completely eliminates large amplitude oscillations from the system, leaving only small amplitude single-well chaotic oscillations feasible, as depicted in Fig. \ref{fig8}(b) for $\xi=0.03$.
		
\begin{figure}
		\centering
		\includegraphics[width=1.0\columnwidth]{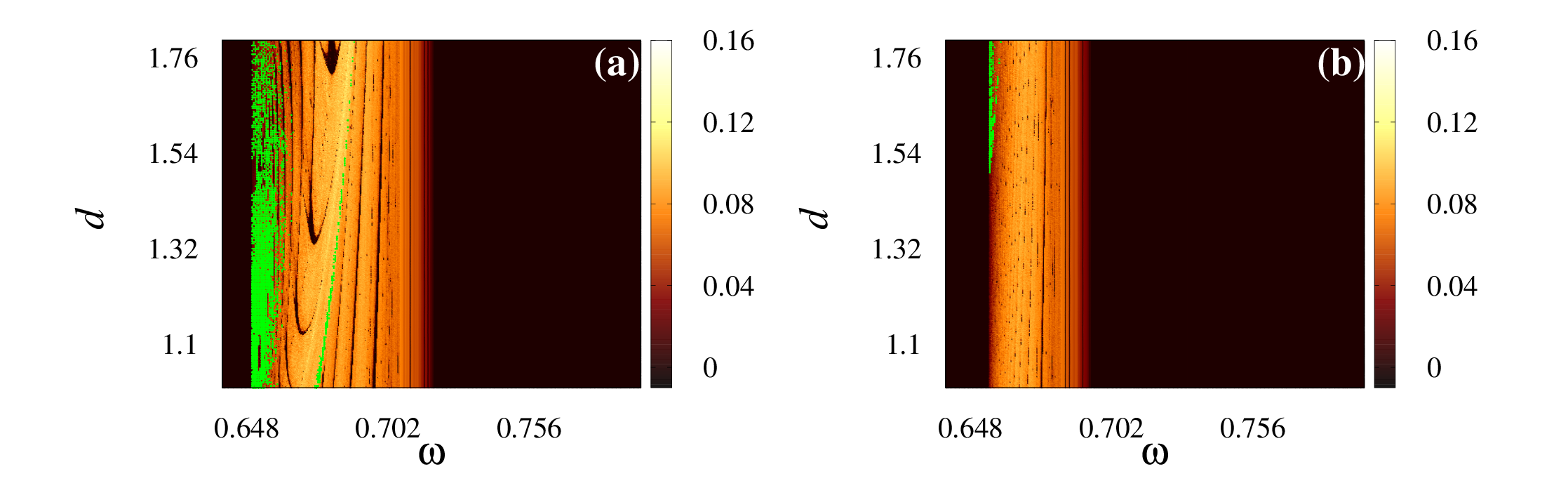}
		\caption{The impact of linear damping on EE is demonstrated. Two parameter phase diagrams of the system (\ref{Eqn1}) within the $\omega-d$ parameter space. Dark brown (gray) regions signify shades negative values in maximum LE values, confirming the regular dynamics of the system. Conversely, positive LE, depicted in remaining color shades, indicate chaotic dynamics. These diagrams showcase the system's dynamics for two distinct values of linear damping strength ($\xi$), exhibit how EE are eliminated as $\xi$ increases. In (a) the phase diagram corresponds to $\xi=0.02$, and (b) is plotted for $\xi=0.03$. The color palettes used match those described in Fig.\ref{fig5}. }
		\label{fig7}
	\end{figure}

		We further extend our analysis to the global dynamics of the system across the $\omega-d$ parameter space, specifically for $\xi=0.02$ and $\xi=0.03$. Our examination for $\xi=0.02$ reveals a mitigation of chaotic dynamics compared to the case with $\xi=0$, leading to a noticeable reduction in the EE region within the parameter space, as shown in Fig. \ref{fig7}(a). Direct comparison with Fig. \ref{fig5} (depicted for $\xi=0$) highlights a significant decrease in the EE region.
		
		Upon increasing the linear damping parameter to $\xi=0.03$, our findings indicate a further contraction of both complex dynamics and the EE region, as supported by Fig. \ref{fig7}(b). Thus, we affirm that the inclusion of linear damping induces the elimination of EE from the system dynamics, leaving only periodic dynamics as a feasible outcome. 
	
	\section{Conclusion}
	\label{sec4}
	In summary of our investigations, we conducted a thorough examination of the dynamics exhibited by the Li\'enard oscillator under the influence of asymmetric potential wells. Compelling evidence has been presented for the occurrence of EE, characterized by sudden, infrequent large-amplitude peaks amid regular small-amplitude oscillations, particularly when modifying the height of the potential well. In the scenario of a symmetric well, the system demonstrates chaotic behavior, with the trajectory oscillating between two potential wells, leading to frequent large bursting oscillations under specific parameter values. However, upon the introduction of asymmetric potential wells, we observed a significant reduction in the frequency of jumping between wells. Consequently, the system trajectory exhibits rare yet recurrent jumps to adjacent wells, giving rise to EE. These observations are substantiated through bifurcation diagrams and Lyapunov exponents. Furthermore, we depict the emergence of EE in the system as a function of external forcing frequency, $\omega$, and the depth of the potential well, $d$ through phase diagram plots. Additionally, we showcase the control of EE by introducing a linear damping term into the system, as validated in our results. 
	
	 Our research not only expands the theoretical understanding of nonlinear dynamics by elucidating the occurrence and characteristics of EE in asymmetric potential wells but also offers practical insights with significant implications. The findings of this study hold particular relevance in the realm of MEMS, where nonlinear behavior is commonplace \cite{giri2021}. Understanding EE in systems with asymmetric potentials is pivotal for optimizing MEMS device design and functionality. Our study also contributes to enhancing the control and stability of mechanical systems, such as improving vibration dampening and optimizing structural configurations.
		
		To further broaden the impact of our findings, we delve into the multiscale analysis of EE, exploring their manifestation and attributes across diverse temporal and spatial scales. This approach provides valuable insights into the intricate interplay between local dynamics and global system behavior, particularly in complex systems. Moreover, the investigation of control strategies to induce or suppress EE in asymmetric potential systems holds promise for various applications across a range of fields.
		
		In essence, our study represents a significant advancement in comprehending the complex dynamical processes associated with EE. By contributing to the body of knowledge in nonlinear dynamics research, our findings reinforce the credibility and importance of this work within the scientific community.
	
	\section*{Acknowledgement}
	 B K acknowledges SASTRA University for providing Teaching Assistantship. The work of RS, and VKC forms a part of a research project sponsored by the SERB-CRG Grant No. CRG/2022/004784. All the authors thank DST, New Delhi, for computational facilities under the DST-FIST program with project number SR/FST/PS-1/2020/135 of the Department of Physics.\\
	
	\noindent {\bf Author Contributions} All the authors contributed equally to the preparation of this manuscript.\\
	
	\noindent {\bf Data Availability Statement} The authors confirm that the data supporting the findings of this study are available within the article.

\end{document}